\documentclass[twocolumn,showpacs,preprintnumbers,amsmath,amssymb]{revtex4}

\usepackage[dvipdfmx]{graphicx, color}
\usepackage{dcolumn}
\usepackage{bm}

\begin{document}

\title{Ultrafast near infrared photoinduced absorption in a multiferroic single crystal of bismuth ferrite 
}

\author{Eiichi Matsubara$^{1,2}$}
\author{Takeshi Mochizuki$^2$}
\author{}%
\author{Masaya Nagai$^2$}%
\author{Toshimitsu Ito$^3$}%
\author{Masaaki Ashida$^2$}%

\affiliation{$^1$Department of Physics, Osaka Dental University, 8-1, Kuzuha-Hanazono, Hirakata, Osaka, 573-1121,Japan,
}%
\affiliation{$^2$Graduate School of Engineering Science, Osaka University, 
1-3, Machikaneyama, Toyonaka, Osaka, 560-8531, Japan,
}%

\affiliation{$^3$Electronics and Photonics Research Institute, National Institute of Advanced Industrial 
Science and Technology (AIST), 1-1-1, Higashi, Tsukuba,
Ibaraki, 305-8562, Japan
}%


\begin{abstract}
We studied the ultrafast third-order optical nonlinearity in a single crystal of 
multiferroic bismuth ferrite (BiFeO$_3$) in the near-infrared range of 0.5--1.0 eV, where the material 
is fundamentally transparent, at room temperature. With pump pulses at 1.55 eV, which is off-resonant 
to the strong inter-band charge transfer (CT) transition, we observed instantaneous transient absorption with pencil-like temporal profile originating from the two-photon CT transition from the oxygen 2$p$ to the iron 3$p$ levels. In contrast, under pumping with 3.10-eV photons, the pencil-like absorption change was not observed but decay profiles showed longer time constants.
Although the two-photon absorption coefficient is estimated to be 1.5 cm/GW, which is ten (hundred) times smaller 
than that of two(one)-dimensional cuprates, it is larger than those of common semiconductors such as ZnSe and GaAs at the 
optical communication wavelength. 
\end{abstract}
\maketitle
\section{Introduction}
A lot of efforts have been devoted for seeking materials for all optical switching, where large nonlinearity, 
quick response time, and operability at room temperature are demanded. Transition metal oxides of Mott Hubbard insulators 
are known to show ultrafast photoinduced phenomena reflecting broadband optical absorption. Especially in an 
one dimensional cuprate of Sr$_2$CuO$_3$, gigantic photoinduced absorption with an ultrafast decay constant 
of 1 ps occurs owing to the two-photon absorption process \cite{ogasawara, ashida, ashida2}. Since the 
wavelength range of probe pulses for the process includes the optical communication wavelength, such kind of 
materials are hopeful candidates for ultrafast all optical switches. So far, one and two dimensional cuprates 
have been well studied, however, there are not so many reports on other transition metal oxides especially 
three dimensional ones. 

Bismuth ferrite (BiFeO$_3$) is one of the  multiferroics which possesses both ferroelectricity (FE) and antiferromagnetism ($T_C \sim$1100 K, $T_N \sim$ 640 K), this material is studied from viewpoints of both fundamental physics and applications \cite{zhao, lebeugle}. 
Optical pump-probe measurements under excitation with 
above bandgap (2.6--2.8 eV) photons have been performed in films and single crystals of BiFeO$_3$, where the 
authors discussed various photoinduced phenomena such as coherent oscillations due to magnon and phonon 
\cite{ruello,doig,chen}, and spectral modulations in connection with the transport property \cite{yamada,sheu}.
As a nonlinear medium, BiFeO$_3$ has been studied in terms of the second-order nonlinearity in studies such as 
terahertz emission through optical rectification \cite{talbayev1} and ultrafast modulation of spontaneous 
polarization \cite{takahashi}, and second harmonic generation spectroscopy \cite{kumar1,ramirez}. However, 
there has been almost no study on the third-order nonlinearity especially around the optical communication 
wavelength in the near infrared region. 

BiFeO$_3$ is also one of the Mott insulators with ultrabroadband optical response continuously ranging from the 
far-infrared (terahertz) to the ultraviolet regions, reflecting strong interactions not only between 
electrons but also between electrons and lattice, spin, orbital degrees of freedoms, so that the material 
makes us expect strong and ultrafast nonlinear optical responses in the near-infrared below-gap region 
as ever reported in low-dimensional cuprates \cite{ogasawara, ashida}. 
Furthermore, BiFeO$_3$ has the aspect of a ligand system, 
in which $d$-$d$ transitions play important role in the optical responses in the visible and the near 
infrared regions, which makes the physics of the material more interesting. 

As for the sample preparation, the fabrication of a single crystal of BiFeO$_3$ has long been difficult. 
Recently, single crystals grown by the flux method have been utilized in many studies 
\cite{lobo, talbayev1, cazayous, singh, rovillain, xu, talbayev2, sheu, pisarev}, however, the obtained 
size of the crystals has been limited. In 2011, Ito \textit{et al.} realized the fabrication of large single 
crystals of BiFeO$_3$ by using a modified floating zone method with laser diodes for heating \cite{ito}. 
Besides the large size (4 mm in diameter), the leakage current under high bias voltage is extremely low. 
Hence, we think this crystal will enable us to extract only the intrinsic optical property of the material 
by making full use of the interaction length. 
Thus, in the present study, we optically pump the large single crystal of BiFeO$_3$ with femtosecond pulses and probe 
the transient absorption change mainly in the near infrared region, to explore unknown ultrafast nonlinear 
response of BiFeO$_3$. 

\section{Experimental}
A large single crystal of BiFeO$_3$ with a diameter of 4 mm was grown with the floating zone method utilizing 
laser diodes as heat sources \cite{ito}. By controlling the oxygen pressure during growth, oxygen 
deficiency is reduced a lot so that the leakage current flowing in the crystal under high bias-voltage 
is extremely low. Also the crystal does not show any magnetic hysteresis in the $M$-$H$ curve which is usually 
observed due to the weak parasitic ferromagnetismin in crystals grown by the flux method. 

The unit cell of BiFeO$_3$ has the rhombohedral ($R3c$) structure, which is essentially a cubic elongated along the [111] direction. The ferroelectric polarization is along this axis. 
BiFeO$_3$ is a G-type antiferromagnet in which all the nearest neighbor spin couplings, including both inter- and intra-plane ones, are antiferromagnetic. Due to the magnetic order, BiFeO$_3$ forms the spiral spin structure with a period of 62 nm along the three equivalent wavevectors of [10-1], [01-1], and [1-10] directions. BiFeO$_3$ is also a charge transfer (CT) type insulator with a bandgap energy of 2.6$-$2.8 eV, which corresponds to the lowest inter-band transition from the oxygen 2$p$ to the iron 3$d$ state. This broad CT band can be further decomposed into several peak structures centered at 2.5, 2.9, 3.2, 4.0, 4.5, and 6.1 eV \cite{pisarev}. This CT band with strong oscillator strengths brings about strong absorption with coefficients of the order of 10$^5$ cm$^{-1}$ \cite{xu}. BiFeO$_3$ also shows relatively weak ($\sim$100 cm$^{-1}$) $d$-$d$ 
absorptions with broad peaks centered at 1.4 and 1.9 eV, which are also split into two by taking the 
spin degrees of freedom into account \cite{ramirez}.

Pump-probe measurements were carried out with a titanium sapphire regenerative amplification system 
(800 nm, 35 fs, 1 kHz) as a light source. A (100)-oriented specimen of BiFeO$_3$ with a thickness of 
170 $\mu$m was pumped with the fundamental (800 nm, 1.55 eV) or the second harmonic pulses (400 nm, 
3.10 eV), and the transient transmission change was probed with signal or idler pulses from an optical 
parametric amplifier. The duration of the pump pulse was measeued to be 60 fs, 
and the cross correlation width of the pump and probe pulses was 100--110 fs, both at the sample. 
The penetration depths given by the inverse of the absorption coefficients \cite{xu} 
are 33  $\mu$m at 1.55 eV, and 0.49 $\mu$m at 3.10 eV, respectively. The near-infrared transmission was measured 
with a combination of a monochromator and a cooled HgCdTe detector. The $1/e^2$ beam sizes of the pump and 
probe pulses were 1.35 mm and 130 $\mu$m in diameter, respectively. First we pumped the specimen with 3.1-eV 
pulses and measured the transient optical conductivity with air-plasma based terahertz time domain 
spectroscopy. As a result, we observed no Drude-like response but 
observed an oscillation of transmission resembling a monocyclic terahertz pulse profile (not shown). 
This result is consistent with the high resistivity confirmed with the dc measurements \cite{ito} and 
the oscillation profile seems to be caused by the generation of terahertz pulses due to the modulation of 
polarization by ultraviolet pulses \cite{takahashi}. 
Since the electrons transport property is determined by oxygen deficiencies, the absence of the transient Drude response proves 
the suppression of them. Such high quality of the crystal will enable us to extract intrinsic property without artifacts.

\section{Results}
Figure 1 shows the time evolutions of the photoinduced absorption change at each probe photon energy (0.5--1.0 eV). 
The pump photon energy is fixed at 1.55 eV. 
At all probe photon energies, we can see steep increase of absorption and its ultrafast recovery. 
The amount of absorption change increases with the probe photon energy. 
The inset shows the expanded profiles around the time origin. Here the temporal profiles, especially for probe-photon 
energies greater than 0.8 eV, characteristically have pencil-like structures with a duration of 100--150 fs. 
This is close to the cross correlation width of the pump and probe pulses (100--110 fs). The decay profiles were well 
fitted by the expression assuming two exponential functions and a constant as, 
\begin{equation}
\Delta\alpha L=A_{0}+A_{1}\textnormal{exp}(-t/\tau_{1})+A_{2}\textnormal{exp}(-t/\tau_{2}).  
\end{equation}
The constant component ($A_0$) certainly has a decay period beyond 1 ns, however, 
it did not exceed 1 ms, which is the separation period of the laser source. The values 
of $\tau_2$ and  $\tau_1$ are fitted to be 50--60 fs and 1--3 ps, respectively. 
In the inset of Fig. 2 (a), we present how the transient absorption profiles probed at 0.9 eV 
are decomposed into three components. Importantly, the dotted curve showing the cross-correlation 
profiles of the pump and the probe pulses coincides with the dominant part of the transient absorption change. 
They appear to be essentially independent of the probe photon energy 
(wavelength) within the accuracy of the measurements. In Fig. 3 (a) 
we present the dependence of the magnitudes of the three components on the pump photon 
density. All the amplitudes of the three components increased with the probe-photon energy. 
More closely, $A_1$ and $A_2$ were proportional to the excitation density, while $A_0$ 
showed the quadratic dependence. We also examined the polarization dependence and 
found the temporal profile did not depend on the polarization of the pump and probe pulses, 
although the magnitude became smaller when the polarizations of two pulses were crossed. 

\begin{figure}[htp]
\centerline{\includegraphics[width=9cm]{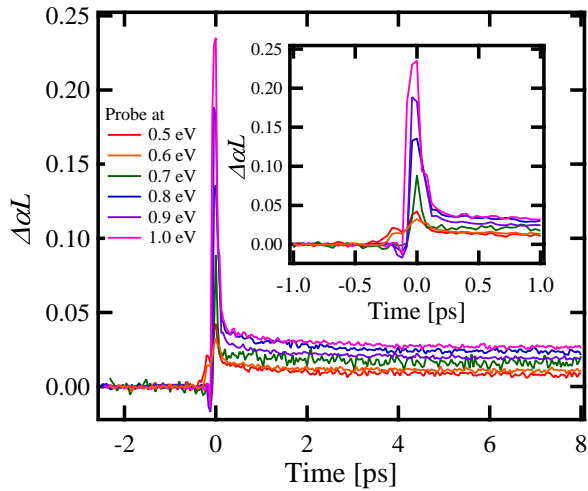}}
\caption{Probe photon energy dependence of the transient absorption profiles $\Delta\alpha L$ in a single crystal of BiFeO$_3$. 
The pump photon energy is fixed at 1.55 eV, while the probe photon energy is varied from 0.5 to 1.0 eV. 
The excitation density is 40 GW/cm$^2$. }
\label{al_probe_energy_dep}
\end{figure}

\begin{figure}[htp]
\includegraphics[width=9cm]{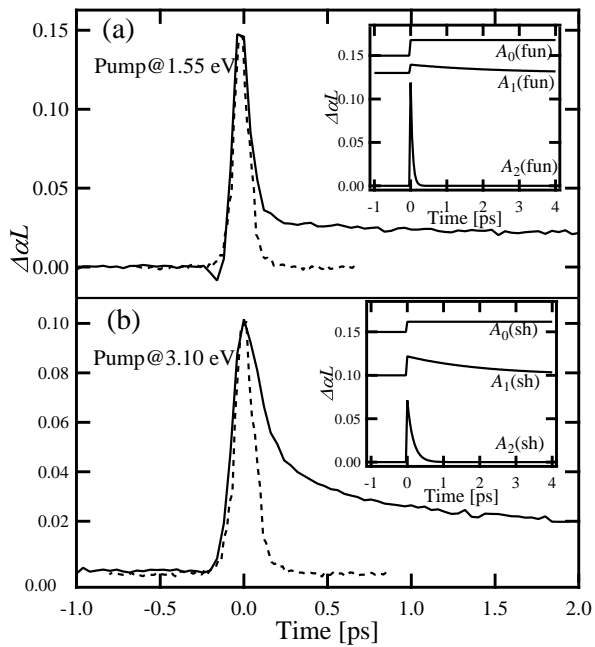}
\caption{Pump photon-energy dependence of the photoinduced absorption $\Delta\alpha L$ at the probe-photon 
energy of 0.9 eV. (a) Temporal profile pumped at 1.55 eV, and (b) that pumped at 3.10 eV. Insets show 
temporal profiles of each ($A_0$, $A_1$, $A_2$) component. The excitation density is 56 GW/cm$^2$. 
Dotted curves show the cross-correlation profiles of the pump and the probe pulses at each pump photon energy. 
Here, the profile at the pump photon energy of 1.55 eV was directly measured, while that at 3.10 eV was 
estimated as the possible maximum one taking into account the group delay dispersion in a BBO crystal and optical fiters 
employed for the measurement. }
\label{fn_vs_sh2}
\end{figure}

To elucidate the physical mechanism, we changed the excitation photon energy to 3.10 eV and 
measured the transient absorption signal at the probe photon energy of 0.9 eV. The profile is 
shown in Fig. 2 (b). The decay curve was also decomposed into three parts with time constants 
of 140--170 fs, 1.9--2.6 ps, and nanoseconds (which can be regarded as constant). Hereafter, 
we denote the fitting results as $A_0$(fun) or $A_0$(sh) to distinguish those pumped by the 
fundamental (1.55 eV) and second harmonic (3.10 eV) pulses. The inset shows the profiles of 
each decomposed decay component. Apparently, the decay time of the fastest component 
for pumping with 3.10-eV pulses ($\tau_2$(sh)) is longer than that with 1.55-eV pulses ($\tau_2$(fun)), 
where the magnitude of $A_2$(sh) is relatively smaller than $A_2$(fun). The profile under pumping 
with 3.10-eV pulses is sharper at the top and does not have the pencil-like structure, which appeared 
under pumping with 1.55-eV photons, although the duration of 3.10-eV pulses is expected to be 
longer than that of the 1.55-eV ones due to the material dispersion in the nonlinear crystal. 
We also examined the dependence on the excitation density as presented with closed markers 
in Fig. 3 (b). $A_2$(sh) was quadratically proportional to the excitation density 
up to 60 GW/cm$^2$, approximately, while $A_0$(sh) and $A_1$(sh) showed linear dependence and 
the slopes become gentle around 40 and 55 GW/cm$^2$, respectively. From these dependences and the absence 
of the pencil-like structure, the origin of photoinduced absorption under pumping with the second harmonic 
(3.10 eV) pulses is different from that with fundamental (1.55 eV) pulses.  


\begin{figure}[htp]
\centerline{\includegraphics[width=9cm]{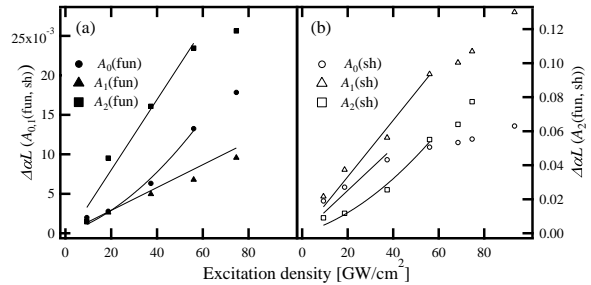}}
\caption{Excitation density dependence of the magnitude of each decay component 
under pumping with (a) fundamental (1.55 eV) pulses , and (b) second harmonic pulses  (3.10 eV) 
at the probe photon-energy of 0.9 eV. 
For clarity, vertical error bars are not indicated, however, the values of error are small so as not to effect the dependence. }
\label{plot_amp_power3}
\end{figure}

\section{Discussions}
\begin{figure}[htp]
\centerline{\includegraphics[width=8cm]{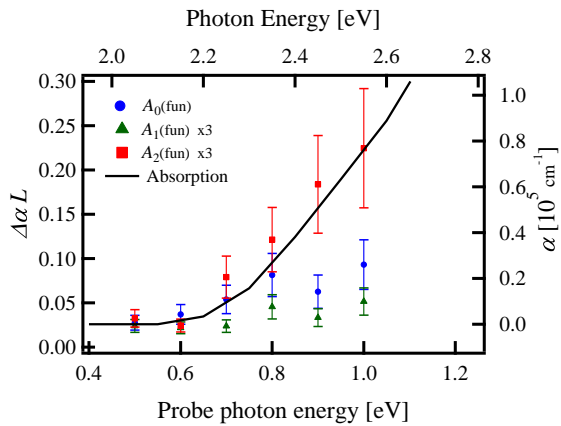}}
\caption{Relationship between the magnitude of decay components and the linear absorption spectrum of BiFeO$_3$. 
$A_0$, $A_1$, and $A_2$ are represented as closed markers as a function of probe photon energy (left and bottom axis). 
Solid curve shows the absorption spectrum (right and top axis) adopted from the data shown in Ref. \cite{pisarev}. }
\label{analysis_spectrum4}
\end{figure}
Now let us elucidate the microscopic origin of the present photoinduced phenomena. First, we focus on 
the pencil-like top structure in the transient absorption profile and the origin of the fastest decay 
component $\tau_1$(fun). Ogasawara \textit{et al.} observed a similar structure in the photoinduced 
transmission change profiles of a quasi-one-dimensional cuprate of Sr$_2$CuO$_3$, which they attributed 
to the two-photon excitation process from the oxygen $2p$ to the even-parity band of Cu which lies at a 
higher frequency than the one-photon CT band \cite{ogasawara, ashida, ashida2}. 
We plot the amplitude of the fastest 
decay component as a function of the probe photon energy as shown in Fig. 4. 
By overlaying the absorption spectrum in such a manner that the horizontal axis (top) is shifted by the pump photon energy of 1.55 eV 
towards higher energy, the magnitude of the fastest component $A_2$(fun) significantly rises up at 0.7 eV, where the 
linear absorption with the corresponding photon energy (2.2--2.3 eV) also starts to increase. From this and 
the pencil-like structure, we conclude that the two-photon absorption process from the O 2$p$ and the 
Fe 3$d$ level gives the fastest decay component as schematically shown in Fig. 5 (a). 
Note the unit cell of BiFeO$_3$ lacks inversion symmetry at temperatures below 1100 K, so that the two-photon 
transition to the one-photon allowed level is allowed. 

Let us further examine the probe photon energy (wavelength) dependence. While $A_0$(fun) significantly increases 
with the probe photon energy, the increase of $A_1$(fun) is minor. According to the result of the pump-probe measurements 
in the ultraviolet range, the total lifetime of the CT excited states is in the order of nanoseconds \cite{sheu}. 
In the present case, one pump photon cannot create the CT excited state, however, the two-photon process can realize it. 
Hence, we consider the $A_0$(fun) component comes from the lifetime of the two-photon excited CT states. We can attribute the gradual increase of $A_0$(fun) with the probe photon-energy to the dependence of the oscillator strength of the CT transition on the photon energy. The pump photon energy of 1.55 eV is also close to the broadband $d$-$d$ transition from $^{6}A_{1g}$ to $^{4}T_{1g}$, so that the $A_1$(fun) component with the decay constant of 1--3 ps probably corresponds to the lifetime of the $d$-$d$ excited states, and the transition to the higher $d$-$d$ levels or upper Hubbard band  induces the transient absorption as schematically shown in Fig. 5 (b).  The relatively small amplitude $A_1$(fun) is 
due to the smaller oscillator strength compared with that of the CT transition by nearly one hundred times. 
As for the channel of the decay, we can think of some process involving the spin system, because 
the $d$-$d$ transitions are known to be accompanied by magnon excitation in the form of sidebands \cite{ramirez}.

Next let us discuss the photoinduced phenomena under pumping with 3.10-eV photons. The time evolution 
basically resembles the ones reported by Sheu \textit{et al.} \cite{sheu}, who pumped a single crystal 
of BiFeO$_3$ with 3.1-eV photons and measured the transient reflection change also at 3.1 eV. 
From the quadratic dependence of $A_2$(sh) on the excitation density, some cooperative interaction 
is certainly involved in the decay process. Similarly to the case in the previous report, the relaxation 
from the CT excited states to the bottom of the band through electron-electron or electron-phonon 
scatterings is highly possible as the origin. As already written, $\tau_2$(sh) slightly decreases with the 
excitation density, which seems to support this speculation. Furthermore, 
while $A_0$(fun) shows the quadratic dependence up to the excitation density of 56 GW/cm$^2$, 
$A_0$ (sh) shows the linear dependence only up to 40 GW/cm$^2$ and then the slope becomes gentle. 
In the case of excitation with fundamental (1.55 eV) pulses, some cooperative interaction seemingly works 
between the two-photon excited CT states. This is the origin of parabolic dependence of $A_0$ on the excitation density. 
In contrast, $A_0$(sh) reflects the one-photon excited CT states, it can easily be saturated at the relatively weak 
excitation density as we observed in Fig. \ref{plot_amp_power3}(b).

Now let us discuss the two-photon absorption coefficient of $\beta$ in BiFeO$_3$. 
Using the transient transmission change of --0.11 at the probe photon energy of 1.0 eV (1.24 $\mu$m), 
$\beta$ is estimated to be 1.5 cm/GW. The $\beta$ in the one-dimensional cuprate of Sr$_2$CuO$_3$ is 
160 cm/GW at the peak of the two-photon band\cite{ogasawara}, while that of the two-dimensional 
Sr$_2$CuO$_2$Cl$_2$ is 12 cm/GW. This dimensionality dependence of the optical nonlinearity in materials 
relating to high $T_C$ cuprate superconductors was closely examined by Ashida \textit{et al.}, in which 
the $\beta$ in the two-dimensional cuprates was typically ten times smaller than that of the one-dimensional 
ones. The mechanism was explained using the cluster model\cite{ashida2}. 
Intuitively, to limit the coherent vibrational oscillations of electrons with large amplitudes in one direction can enhance the optical nonlinearity. While in the case of two or three dimensional materials, motion of electrons spreads into multiple directions 
even if electrons and lattice are vibrated in one direction by linearly polarized optical pulses. Thus the low-dimensional cuprates may be more beneficial in terms of application to optical device. However, the $\beta$ value is still larger than those of common semiconductors such as ZnSe and GaAs at the 
optical communication wavelength \cite{hutchings}. BiFeO$_3$ also has the strong aspect as a multiferroic, multifunctional 
responses relating to the magneto-optic effect can be expected in low temperature or polarization sensitive measurements. 
To explore such phenomena is our future work.


\begin{figure}[htp]
\centerline{\includegraphics[width=9cm]{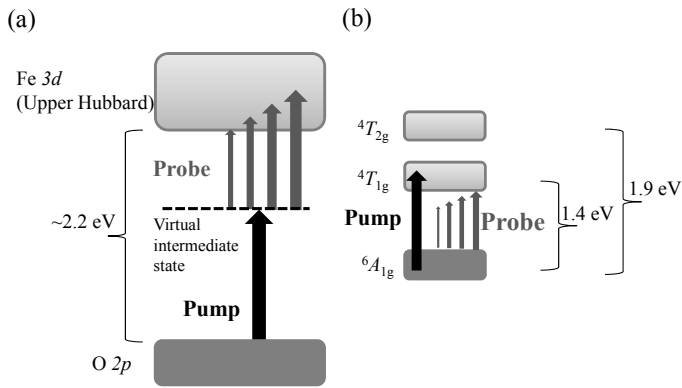}}
\caption{Schematic drawings of the electronic states of bismuth ferrite to show the origin of the ultrafast decay components under 
pumping with fundamental (1.55 eV) pulses. 
(a) Upper Hubbard band of Fe 3$d$ and O 2$p$ level, in connection with the fastest ($A_2$(fun)) 
and the slowest $A_0$(fun) decay components. 
(b) Relationship between the $d$-$d$ levels and the second fastest ($A_1$(fun)) component. }
\label{levels2_2}
\end{figure}
Finally, let us give a brief comment on the relationship between the present results and the optical 
anisotropy. BiFeO$_3$ shows uniaxial birefringence with an extraordinary axis along the 
[111] direction. However, we observed no significant polarization dependence in the results of the 
pump probe spectroscopy. 
We can understand the reason considering the three dimensional chemical bonding structure in BiFeO$_3$; in the present phenomenon, the CT excitation and $d$-$d$ transition 
play important roles, while absorption due to phonons and magnons shows significant polarization dependence. 
At low temperatures, the coherence of elementary excitation will grow up enough to defeat dephasing 
factors and we will observe the transient anisotropic optical response originate from coherent phonon 
and magnon.

\section{Conclusions}
In conclusion we demonstrated the ultrafast photoinduced absorption spectroscopy in a single crystal of BiFeO$_3$ using sub-100-fs optical pulses in the near-infrared and ultraviolet ranges. With pump pulses at 1.55 eV, we observed sharp decrease in transmission which remained constant around the minimum for approximately 100 fs, it decayed with a time constant as short as 50--60 fs. Examining the dependences of the signal intensity on pump fluence and probe photon energy, we conclude the dominant transient absorption originates from the two-photon absorption process from the oxygen 2$p$ to the iron 3$d$ level. The decay profile also has two other components with constants of 1--3 ps, and nanoseconds which can be fitted as a constant in the present observation range. We attribute them as some relaxation process involving spins, and the total lifetime of the CT excited electrons via the two-photon process. The two-photon absorption coefficient is estimated to be 1.5 cm/GW, which is ten (hundred) times smaller than that of two(one)-dimensional cuprates. This result shows such ultrafast response is common in strongly correlated systems and may reflect the dimension dependence of the nonlinearity. The present $\beta$ value is still larger than those of common  semiconductors such as ZnSe and GaAs at the optical communication wavelength. With pump pulses at 3.10 eV, the pencil like two-photon absorption did not appear, but the profiles had longer time components of 140--170 fs, 1.9--2.6 ps, and nanoseconds. We attribute them to the collective electron-electron and electron-phonon scattering in the CT band, some decay process involving spin or band to band relaxation, and the total lifetime of the CT excited levels, respectively.

\bibliography{apssamp}

\end{document}